\documentclass[structabstract]{aa}  
\usepackage{longtable}
\usepackage{latexsym}
\usepackage{amssymb}
\usepackage{lscape}
\usepackage{graphicx}
\usepackage{txfonts}
\title{H$\alpha3$: an H$\alpha$ imaging survey of HI selected galaxies 
from ALFALFA\thanks{Observations 
taken at the  observatory of San Pedro Martir (Baja California, Mexico), belonging to the
Mexican Observatorio Astron\'omico Nacional.}
}

\subtitle{I: Catalogue in the Local Supercluster}

\author{G. Gavazzi \inst{1}
\and M. Fumagalli \inst{2}
\and V. Galardo \inst{1}
\and F. Grossetti \inst{1}
\and A. Boselli \inst{3}
\and R. Giovanelli \inst{4}
\and M.P. Haynes \inst{4}
\and S. Fabello \inst{5}
}

\authorrunning{G. Gavazzi et al.}
\titlerunning{H$\alpha3$: H$\alpha$ imaging survey of HI selected galaxies from ALFALFA}

\institute{Universit\`a degli Studi di Milano-Bicocca, Piazza della Scienza 3, 20126 Milano, Italy\\
\email {giuseppe.gavazzi@mib.infn.it}
\and
Department of Astronomy and Astrophysics, University of California, 1156 High Street, Santa Cruz, CA 95064, USA\\
\email {mfumagalli@ucolick.org}
\and
Laboratoire d'Astrophysique de Marseille, UMR 6110 CNRS, 38 rue F. Joliot-Curie, F-13388, Marseille, France\\
\email {alessandro.boselli@oamp.fr}
\and
Center for Radiophysics and Space Research, Space Science Building, Ithaca, NY, 14853\\
\email {haynes@astro.cornell.edu, riccardo@astro.cornell.edu}
\and
Max Planck Institute for Astrophysics, Garching, Karl-Schwarzschild-Str. 1, 
Postfach 1317, D-85741 Garching, Germany\\
\email {fabello@mpa-garching.mpg.de}
}

\begin{document}
\date{Received December 15, 2009; accepted}

 
  \abstract
      {
    We present H$\alpha3$ (acronym for H$\alpha-\alpha\alpha$), an H$\alpha$ 
    narrow-band imaging survey of $\sim 400$ 
    galaxies selected from the HI Arecibo Legacy Fast ALFA Survey (ALFALFA) 
    in the Local Supercluster, including the Virgo cluster.} 
  {By using hydrogen recombination lines as a tracer of recent star formation, we aim to 
    investigate the relationships between atomic neutral gas and newly formed stars
    in different environments (cluster and field), morphological types 
    (spirals and dwarfs), and over a wide range of stellar masses 
    ($\sim 10^{7.5}-10^{11.5}$ M$_\odot$).}
  {We image in H$\alpha$+[NII] all the galaxies that contain more than $10^{7}$ 
    M$_\odot$ of neutral atomic hydrogen
   in the sky region  
    $\rm 11^h < R.A. <16^h\, ; \,4^o< Dec. <16^o; 350<$cz$<2000$ $\rm km ~s^{-1}$
     using 
    the San Pedro Martir 2m telescope. This survey provides a complete census 
    of the star formation in HI rich galaxies of the local 
    universe.}
  {We present the properties of the galaxy sample, 
    together with H$\alpha$ fluxes and equivalent widths. 
    We find an  excellent agreement between the fluxes determined from our images
    in apertures of 3 arcsec diameter and the fluxes derived from the SDSS 
    spectral database. 
    From the H$\alpha$ fluxes corrected for galactic and internal extinction and for
    [NII] contamination we derive the global star formation rates (SFRs). 
  }
  {}
      

   \maketitle

%

\section{Introduction}

The large amount of multi-wavelength data from recent and ongoing
surveys is providing a wealth of information on the different phases
of the interstellar medium (ISM), the stellar content and the 
present day star formation rates (SFRs) in nearby galaxies. Complemented with
results from numerical simulations and theory, these observations
contribute to our understanding of the basic process that regulates
the life of a galaxy: the conversion of gas into stars.
However, crucial questions remain open on which gas phase (on which scale) is ultimately
responsible for new star formation, on which tracers for
the SFR are unbiased, and on the role of very massive stars and of the environment 
in shaping what is the observed luminosity in local galaxies.

Half a century has passed since Schmidt (1959) discovered a foundamental relation between 
the surface density of the star formation and that of the gaseous component in 
galaxies\footnote{The 50th anniversary from the original Schmidt (1959) paper 
was celebrated during the conference SFR@50 held in Spineto in 2009}, today known 
as the Kennicutt-Schmidt (KS) law (Schmidt 1959, Kennicutt 1989, 1998). 
Since then, a large number of theoretical and observational studies have addressed 
the origin of this correlation. Modern observations reveal a relation  
between molecular gas and star formation rate surface density 
(Wong \& Blitz 2002, Kennicutt et al. 2007, Bigiel et al. 2008)
within the optical radius
where CO seems to be a reliable tracer of molecular hydrogen.
While this contrasts the original form of the KS law 
in which a correlation between the SFR and the much more extended atomic gas 
is evident over several order of magnitude in SFR,
this finding is consistent with the basic picture of star formation in giant molecular 
clouds. But it is unclear whether molecular hydrogen drives this correlation (Krumholz et al. 2011, Glover \& Clark 2011)
and departures from this observed relation are still a matter of debate (Fumagalli \& Gavazzi 2008,
Bigiel et al. 2010, Schruba et al. 2011).

Whereas there is unanimous consent that high luminosity late-type galaxies display low 
specific star formation rate (SSFR), i.e. SFR per unit stellar mass, 
as expected from  \emph{downsizing}  (e.g. Gavazzi et al. 1996), the behavior of 
dwarf galaxies, whose SSFR spans a range exceeding two orders of magnitude (Lee et al. 2007), 
is poorly understood. In addition, the SFR inferred from the H$\alpha$ hydrogen recombination
line in these systems or in the outskirt of disks, systematically underpredicts the one 
derived from the UV light  (Meurer et al. 2009, Lee et al. 2009) to the point that 
doubts have been cast on the universality of the initial mass function (IMF; e.g. Meurer et al. 2009) 
and on the reliability of hydrogen recombination lines to trace star formation (Pflamm-Altenburg et al. 2007).
However, uncertainties in the dust extinction  (Boselli et al. 2009),
star formation history (Weisz et al. 2011)  and a stochastic star formation rate 
(Fumagalli et al. 2011) can explain equally well the observed luminosity even for a universal IMF.

\begin{figure*}[!t]
\centering
\includegraphics[width=19cm]{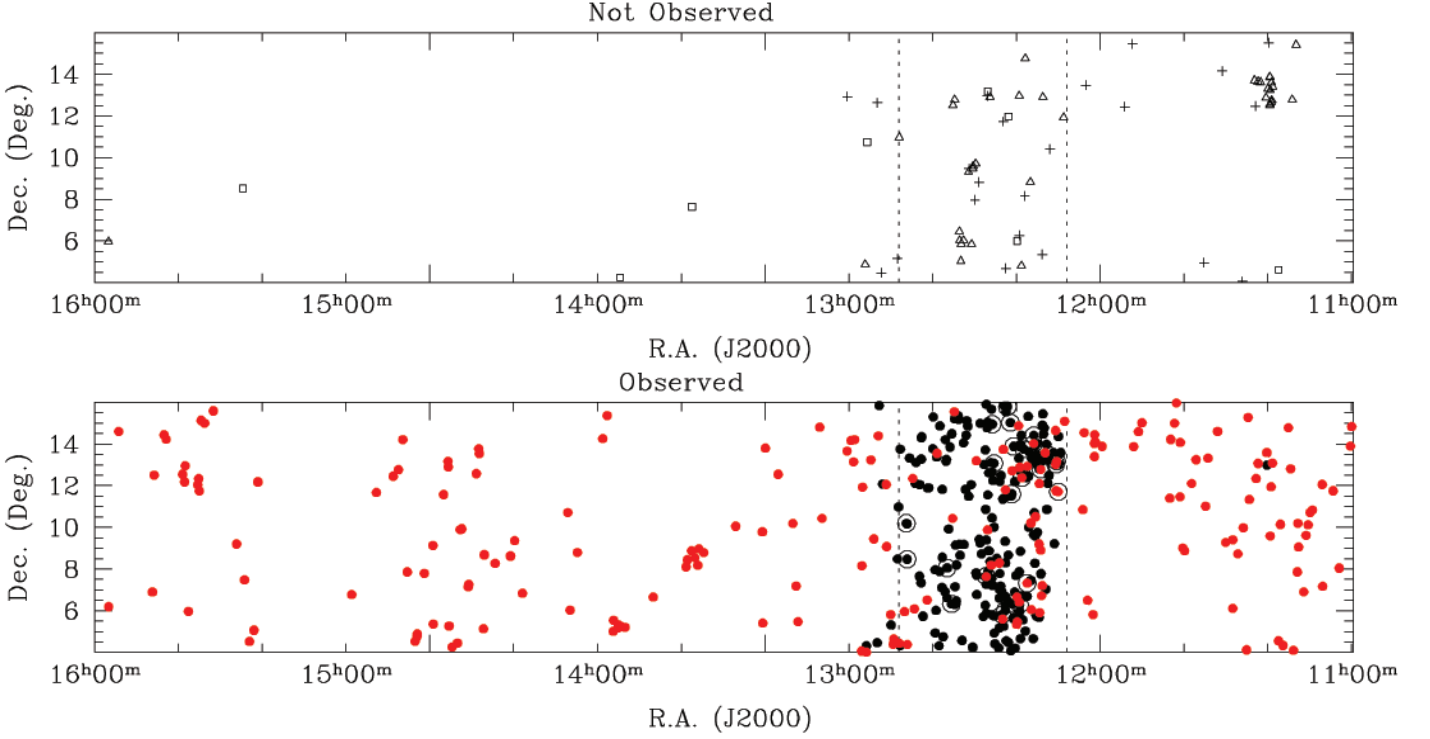}
 \caption{Bottom panel. Sky distribution of 409 HI selected galaxies observed in the present survey, 383 with $F_{HI} > 0.7 {\rm~Jy~km~s^{-1}}$ (filled circles) 
 and 26 with $<$ 0.7 $\rm~Jy~km~s^{-1}$ (big empty circles). 
Red symbols refer to 233 new sources observed in 2006-2009 whose fluxes are presented 
in this paper. Top panel. 68 HI targets that matches our selection criteria 
but that were not observed because: 8 lie too close to bright stars; 
38 are either debris of ram pressure stripped gas or their associated galaxy is too faint to be seen on 
SDSS plates (triangles);  20 that will be consider in future runs (crosses). 
The two vertical broken lines mark the adopted boundaries of the Virgo cluster.}
\label{campione}
\end{figure*}

Similarly, the debate on the role of the environment in shaping the star 
formation properties of galaxies is still open (see a review by Boselli \& Gavazzi 2006). While 
it is observed that atomic (e.g. Gavazzi et al. 2002b, Cortese \& Hughes 2009, Rose et al. 2010)  and, in highly perturbed systems, molecular 
(Vollmer et al. 2008, Fumagalli et al.2009, Vollmer et al. 2009) gas depletion result in a low level of star formation, simulations of 
ram-pressure stripping have suggested different degrees of enhancement in the SFR of perturbed 
galaxies (e.g. Kronberger et al. 2008, Kapferer et al. 2009, Tonnesen \& Bryan 2009). And studies of the H$\alpha$ morphology in galaxies 
within rich groups of clusters show a mix of global suppression and truncation of the H$\alpha$ disks
(Vogt et al. 2004, Koopmann \& Kenney 2004, Fumagalli \& Gavazzi 2008, 
Welikala et al. 2008, Rose et al. 2010), but a definitive assessment of the relative importance
of these different perturbations is still lacking.

To address some of these open issues, we have recently completed 
H$\alpha3$, an H$\alpha$ narrow-band imaging survey at the 2.1m telescope of 
the San Pedro Martir (SPM) Observatory. Our sample includes $\sim 400$ 
nearby galaxies, selected from the ongoing blind HI Arecibo Legacy Fast ALFA Survey 
(ALFALFA; Giovanelli et al. 2005) in the Spring sky of the Local Supercluster, including 
the Virgo cluster, in the velocity window $350<cz<2000$ $\rm km~s^{-1}$.
Together with ancillary multifrequency data and complemented 
by similar surveys  (Meurer et al. 2006) or with optically selected samples (James et al. 2004, Kent et al. 2008), 
these observations provide a complete census 
of the SFR in the local universe traced by hydrogen recombination lines
(see also Bothwell et al. 2009).

The present paper, first of a series, presents the basic properties of the
H$\alpha3$ dataset in the Local Supercluster (Sect. \ref{sample}).  
After a description of observations (Sect. \ref{observ}) and data 
reduction (Sect. \ref{data}), we list  
previously unpublished H$\alpha$ fluxes and equivalent widths for 235 galaxies. 
Summary and future prospects follow
in Sect. \ref{summary}, while in Appendix \ref{atlas} we give an Atlas of the images of the sampled galaxies.\\
Paper II of this series will contain the analysis of the integrated 
quantities (global SFR) produced by H$\alpha3$ in the Local Supercluster, and will
      investigate the relationships between atomic neutral gas and newly formed stars
      in different environments (cluster and field), morphological types 
      (spirals and dwarfs), and over a wide range of stellar masses 
      ($\sim 10^{7.5}-10^{11.5}$ M$_\odot$). \\
  Paper III will contain the extension of H$\alpha3$ to the Coma Supercluster
  ($\rm 10^h < R.A. <16^h\, ; \,24^o< Dec. <28^o; 3900<cz<9000$ $\rm km ~s^{-1}$).\\
  The analysis of the H$\alpha$ morphology from H$\alpha3$ in both the Local and the Coma Superclusters
  will be carried out in Paper IV, which will address the comparison of the 
  effective radii at H$\alpha$ and $r$ band as a function of morphological type, and
  the determination of other structural parameters such as the Concentration index, the Asymmetry and the Clumpiness
  parameters introduced by Conselice (2003).
  Throughout the paper we adopt $H_0=73 \rm ~km~s^{-1}~Mpc^{-1}$.

\section{The Sample}\label{sample}

\subsection{Selection}

Our sample is drawn from the 900 square degree region 
$\rm 11^h <R.A. <16^h\, ; \,4^o< Dec. <16^o$; $350<cz<2000\rm ~km~s^{-1}$, 
covering the Local Supercluster, including the Virgo cluster\footnote{The 
lower velocity limit is dictated by the fact that none of interferometric filters
available at SPM covers the H$\alpha$ line for redshift $<$ 350 $\rm km ~s^{-1}$. Furthermore
some galaxies have been serendipitously observed in spite of having $cz > 2000~{\rm km~s^{-1}}$, 
but they do not constitute a complete sample. In the Virgo cluster however we extend the 
velocity coverage of H$\alpha3$ to $\rm 350<cz<3000~{\rm km~s^{-1}}$ in order to include the 
finger of God of the cluster in its full extent (Gavazzi et al. 1999).}. This region has been fully mapped 
by ALFALFA\ that is providing us with a complete sample of HI selected galaxies, 
with masses as low as $10^{6.5-7.0} ~M_\odot$, 7.7 times deeper than HIPASS, the HI Parkes All-Sky 
Survey  (Meyer et al. 2004)\footnote{
 As introduced in Giovanelli et al. (2005)  ALFALFA is
  a noise-limited survey rather than a flux-limited one. At any given integrated $\rm H_I$  mass 
  the 21 cm flux per velocity channel is inversely proportional to the width of the $\rm H_I$  profile, thus
  to the galaxy inclination.  The completeness and sensitivity of ALFALFA are well understood and discussed
  in detail in  Saintonge (2007), Martin at al. (2010) and Haynes et al. (2011).}. 
The most comprehensive catalogue containing 40\% of the entire ALFALFA coverage
is given in   Haynes et al. (2011), superseding previous ALFALFA publications (e.g. Giovanelli et al. 2007)
for region $\rm 11^h44^m <R.A. <14^h00^m\, ; \,12^o< Dec. <16^o$, and Kent et al. (2008)
for region $\rm 11^h26^m <R.A. <13^h52^m\, ;\,4^o< Dec. <12^o$).

The goal of the H$\alpha3$ survey is to follow up with 
H$\alpha$ imaging observations the ALFALFA\ targets with high S/N (typically S/N$~>6.5$)
and good match between two independent passes  (code = 1 sources; Giovanelli et al. 2005 and Haynes et al. 2011).
In addition, we limit to objects with HI fluxes $F_{\rm HI} >  0.7 {\rm~Jy~km~s^{-1}}$.
At the distance of 17 Mpc, assumed for the Virgo cluster, a flux limit  $F_{\rm HI}=0.7 {\rm~Jy~km~s^{-1}}$ 
corresponds to a HI mass $M_{\rm HI}=10^{7.7} ~{\rm M_\odot}$.

Figure \ref{campione} illustrates the sky region under study. The bottom panel contains the 383
sources that have been observed in H$\alpha$. 
In addition, 26 sources in the Virgo cluster (big open circles)
were observed in spite of having $F_{\rm HI} < 0.7 {\rm~Jy~km~s^{-1}}$
and hence not strictly belonging to the H$\alpha3$ sample selection.
This brings the total number of observed galaxies to 409.

\subsection{Completeness}
\label{compl}

\begin{figure}[!t]
\centering
\includegraphics[width=9cm,height=9cm]{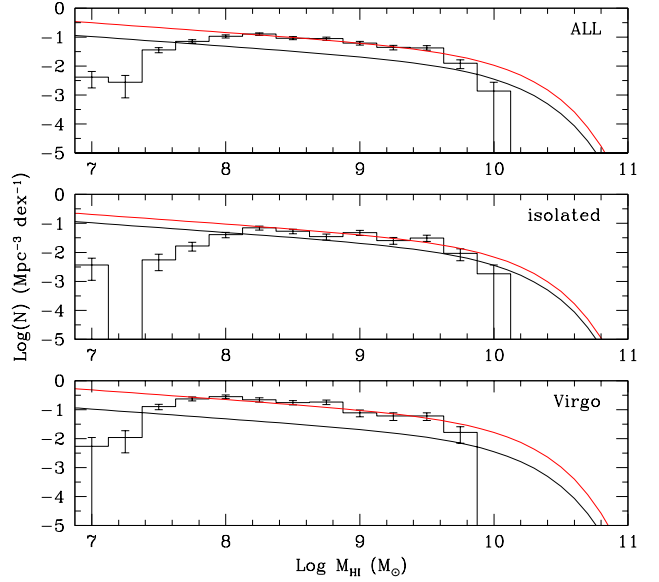}
 \caption{Comparison of the observed HI mass distribution in the three subsamples (histograms) with the HI mass function
 of Martin et al. (2010) (black lines) and one corrected for the overdensity in the Local Supercluster (red lines). 
 }
\label{comp_HI}
\end{figure}

The top panel of Figure \ref{campione} shows 68 ALFALFA\ sources that have not been observed
by H$\alpha3$. 
Among them, 38 (triangles) do not have any optical counterpart because they are either debris 
of ram pressure stripped gas (Kent et al. 2007) or too
faint in optical light to be visible on SDSS plates.
These would have been undetected for the typical 
exposure time of our survey (see Sect. \ref{data}). Of the remaining, 8 galaxies (squares) 
are close to bright stars whose charge bleeding would have destroyed 
the requested photometric accuracy. 
Finally, 20 galaxies will be considered in future runs (crosses).
After accounting for these missing sources, the completeness yet achieved is 
87\% in Virgo and 82\% outside, normalized to the ALFALFA\ catalogue.
To further investigate the HI completeness of H$\alpha3$, i.e. the limiting HI mass 
above which H$\alpha3$ is complete,
we compare in Fig. \ref{comp_HI} the ALFALFA HI mass function 
given by Martin et al. (2010) (black line),
representative of the whole Local Universe sampled by ALFALFA, 
which is well represented by a Schechter function with $\alpha$=-1.33, $\Phi_*=4.8\cdot 10^{-3} \rm Mpc^{-3} dex^{-1}$, 
$M_*=10^{9.96}$\footnote{A more thorough derivation of the ALFALFA HI mass function is in Hallenbeck et al. in preparation.}
with the HI Mass distribution in H$\alpha3$ (histogram). The red curve is the ALFALFA HI mass function 
whose $\Phi_*$ has been normalized to the volumes sampled by  H$\alpha3$, separately for Virgo 
and the isolated volume, to account for the overdensity
in the two subsample with respect to ALFALFA. 
This has been done by dividing the integral of the ALFALFA HI mass function 
in the interval $10^{8-9.75}$ $M_\odot$ by the integral of the observed histogram in the same interval.
The normalization coefficients are 1.96 (isolated), 2.99 (all), 6.01 (Virgo).
The agreement between the red line and the histogram is 
very satisfactory above log $M_{HI}=8$ $M_\odot$, that can be assumed as the HI completeness limit of H$\alpha3$.
The data and the red line depart above log $M_{HI}=9.75$ $M_\odot$ due to cosmic variance as the number of high HI mass galaxies 
sampled by H$\alpha3$ is limited. This is an effect of HI deficiency and the highest HI mass objects are too rare to be found
in the small volume sampled by H$\alpha3$. 
The lack of objects with log $M_{HI}>9.75$ $M_\odot$ is however only of 1-2 objects.\\
The optical completeness of H$\alpha3$ cannot be determined as accurately as for the HI mass because 
the optical luminosity function of the HI selected galaxies is unknown. However
we empirically determine the optical completeness by deriving the cumulative distribution  in 0.5 mag 
bins of $i$-band luminosity of
galaxies in the three observed volumes: isolated, Virgo, all. The cumulative distribution flattens at
$M_i>-15.25$, (corresponding to log $(M_{lim}/M_\odot)=7.8$). This represents the $i$-band 
completeness limit of our HI selected sample that is largely composed of late-type galaxies.

\subsection{Ancillary Data}
\label{ancillary}

The whole region of interest has been covered with imaging and spectroscopic observations by the 
SDSS (DR7, Abazajian et al. 2009). 
However, given the proximity of the surveyed galaxies, their angular size often exceeds several arcmin, making
the shredding problem (Blanton et al. 2005a,b,c) that affects the 
pipeline of SDSS particularly severe.
For this reason and owing to fiber conflict that reduces the number of galaxies with nuclear spectra, 
the SDSS spectral database is not fully complete/reliable for the nearby Universe 
($z<<0.2$). As for example, the catalogued magnitudes are often wrong by several magitudes.
To avoid this problem, the individual $g$ and $i$ band
SDSS images centered on each galaxy 
targeted  by  H$\alpha3$ were downloaded from the SDSS archive. Many of the large galaxies are cut in several pieces
belonging to adjacent "tiles". These images were downloaded individually and combined to cover a
sufficient area to contain
the whole galaxies as well as some empty sky. The background was estimated and subtracted 
using the tasks MARKSKY and SKYSUB in IRAF\footnote{IRAF is the Image Analysis and
Reduction Facility made available to the astronomical community by the
National Optical Astronomy Observatories, which are operated by AURA,
Inc., under contract with the U.S. National Science Foundation. STSDAS
is distributed by the Space Telescope Science Institute, which is
operated by the Association of Universities for Research in Astronomy
(AURA), Inc., under NASA contract NAS 5--26555.} GALPHOT\footnote{developed for IRAF - STSDAS mainly by W. Freudling, J. Salzer, and
M.P. Haynes (Haynes et  al. 1999) and adapted by L. Cortese and S. Zibetti to handle H$\alpha$ data.}. 
The background subtracted frames were inspected individually and background objects and foreground stars 
were masked when found within or near the galaxies of interest. 
The photometry in the edited frames was obtained using $QPHOT$ 
in IRAF  by integrating the counts 
within a circular aperture (determined in $i$-band) containing all the flux. This procedure provides an accurate estimate of the total
$g$ and $i$ magnitudes (see Table \ref{basic1}). During this process the major and minor diameters of the galaxies were 
crudely determined using elliptical regions adapted to the shape of galaxies (see Table \ref{basic1}) in DS9.

The distance to the galaxies belonging to the Virgo cluster is computed according to
Gavazzi et al (1999), who adopts 17 Mpc for members to Virgo A, and for the N, S, E clouds,
23 Mpc for Virgo B (M49), and 32 Mpc for clouds M and W.
These values are consistent with the more modern determination obtained with the surface brightness fluctuation
method on HST-ACS images by Mei at al. (2007).
For all other members of the Local Supercluster we adopt the galactocentric (GSR) distances listed in NED.

The HI mass is computed using 
$M_{HI}= 2.36 \cdot 10^5 \cdot  S_{21} \cdot D^2$, 
where D is the distance to the source in Mpc and S$_{21}$  is the integrated flux under the HI profile in units of Jy-km s$^{-1}$
from ALFALFA.

The stellar mass is derived from the $i$ magnitudes and $g-i$ color using the
Bell's et al. (2003) recipe:
log $M_{star}= -0.152+0.518 \cdot (g-i)$ + log $i_{lum} M_{\odot}$, 
where $i_{lum}$ is the $i$ band luminosity in solar units.

\subsection{Optical Properties}

 For a large majority 
of the 224 Virgo galaxies ($\rm 12^h05^m <R.A. <12^h50^m\, ; \,4^o< Dec. <16^o$; 
$cz < 3000~{\rm km~s^{-1}}$; black filled circles in Figure \ref{campione}) 
the H$\alpha$ data have been already published 
in previous papers (Gavazzi et al. 2002a; Boselli \& Gavazzi  2002; 
Boselli et al. 2002; Gavazzi et al. 2002b; Gavazzi et al. 2006). Images and fluxes are 
publicly available via the GOLDMine web server (Gavazzi et al., 2003).

Properties for the 235 unpublished sources observed in the period 2006-2009 (red filled circles 
in Fig. \ref{campione}) are presented in  Table \ref{basic1}.
Individual entries are as it follows:
\begin{itemize}
\item{Column 1:} AGC designation, from Haynes et al. (2011). AGC numbers coincide with UGC numbers
     for those galaxies included in the UGC (Nilson 1973); 
\item{Columns 2 and 3:} optical celestial coordinates (J2000);
\item{Columns 4, 5, 6 and 7:} CGCG (Zwicky et al. 1968), UGC (Nilson 1973), NGC (Dreyer 1888) and IC (Dreyer 1908) designations;
\item{Column 8:} morphological type, from NED or classified by the authors on   SDSS $i$-band images;
\item{Column 9:} heliocentric velocity of the HI source, c$z_\odot$ in km s$^{-1}$ from Haynes et al. (2011); 
\item{Columns 10 and 11:} major an minor optical diameters from NED or measured
with ellipses on SDSS $i$-band frames (see section \ref{ancillary}). These are consistent with
$25^{th}$ $\rm mag arcsec^{-1}$ isophotal diameters;
\item{Columns 12 and 13:} $i$ and $g$ integrated (AB) magnitude obtained on SDSS frames (see section \ref{ancillary}).
\item{Columns 14: adopted galactocentric (GSR) distances as given by NED (Mpc)}.
\end{itemize}

\begin{figure*} 
 \centering
\begin{tabular}{c c}
\includegraphics[scale=0.35]{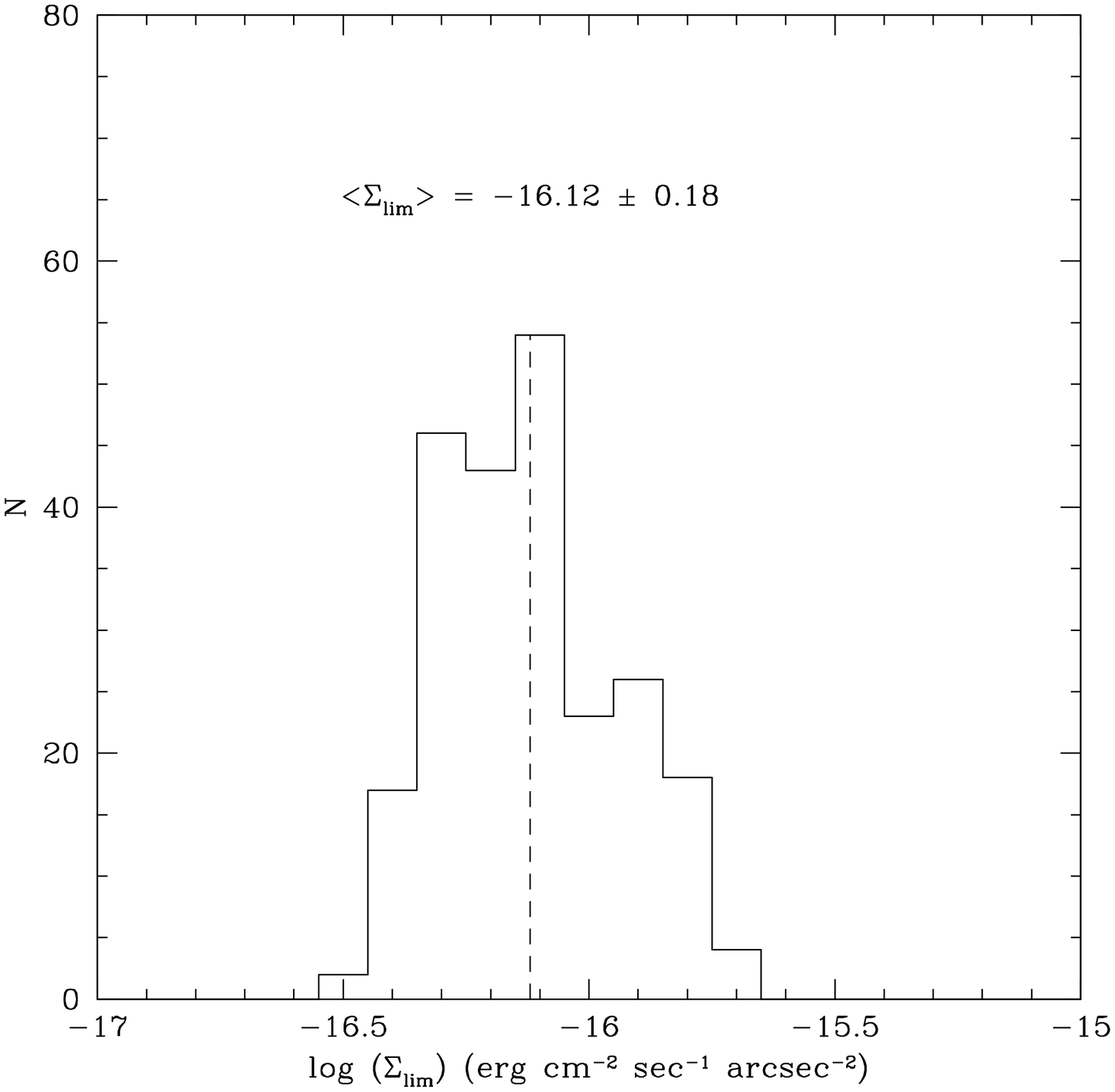}&\includegraphics[scale=0.35]{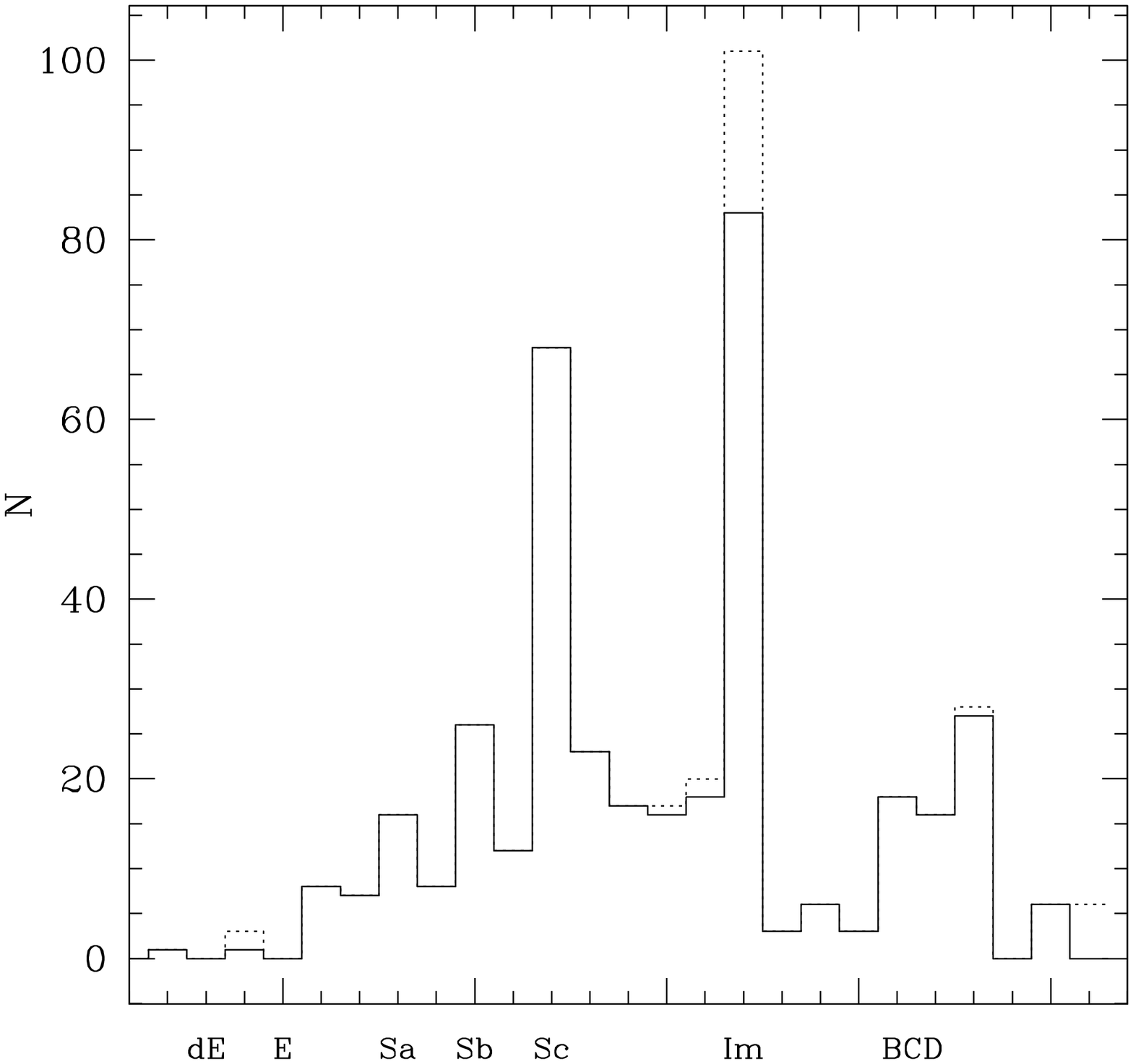}\\
\includegraphics[scale=0.35]{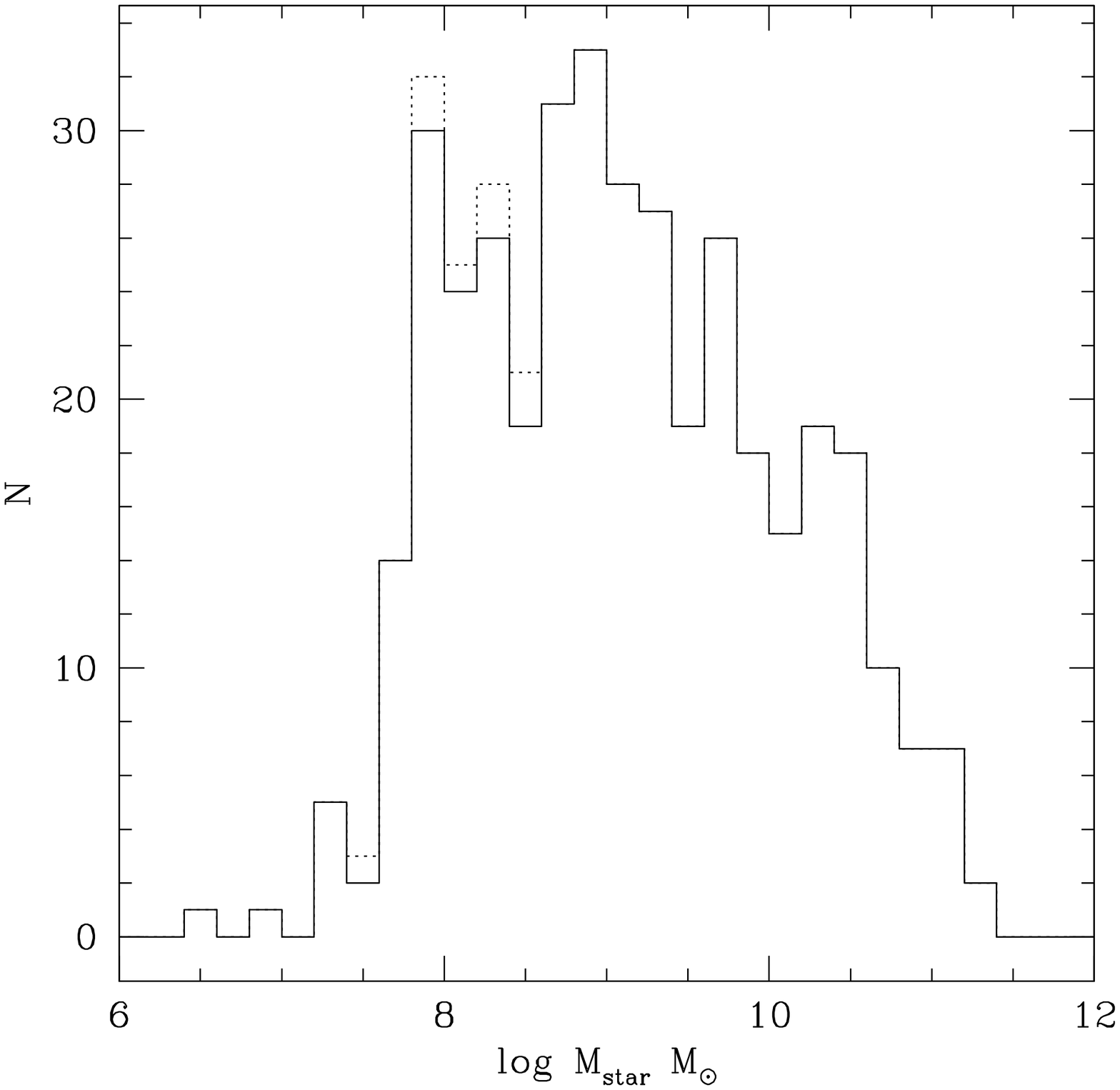}&\includegraphics[scale=0.35]{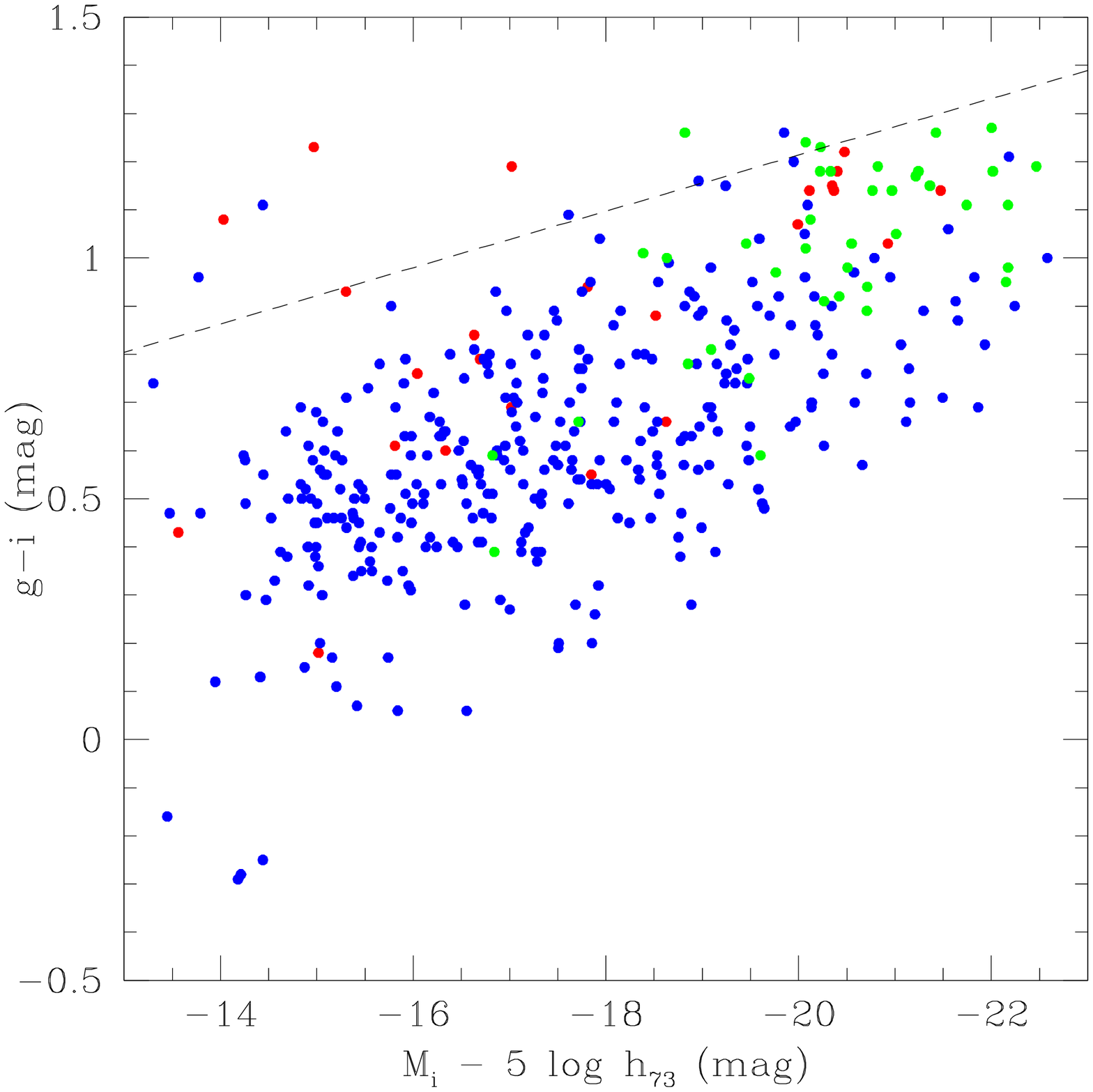}\\
\end{tabular}
\caption{Properties of the H$\alpha3$ sample, compared against the entire ALFALFA\ catalogue 
(dotted lines) and the one restricted to galaxies with optical counterparts (dashed lines). 
Panel (a): 1$\sigma$ limiting surface brightness ($\rm erg~cm^{-2}~sec^{-1}~arcsec^{-2}$) in the H$\alpha$ NET images.
Panel (b): Morphological types. Panel (c): Stellar masses from $i$-band photometry.
Panel (d): Color ($g-i$) magnitude ($i$ band) diagram (color coded by morphology: red=early, blue=disk; green=bulge+disk)
(SDSS magnitudes are uncorrected for internal extinction).
H$\alpha3$ is a homogeneous survey, complete down to a SFR density of $3x10^{-9}$ M$_\odot$ yr$^{-1}$ pc$^{-2}$ 
(1 $\sigma$)
and HI~ masses of $10^8$ M$_\odot$. This sample spans a wide range in color, morphological type,
colors and stellar masses that allows a comparison of the SFR over a broad parameter space.
}
\label{sampleprop}
\end{figure*}

An overview of the sample properties is given in Figure \ref{sampleprop}. 
Panel (a) shows the limiting H$\alpha$ fluxes, computed from the pixel to pixel 1$\sigma$ sky fluctuation. 
We note that most of the galaxies lie in a quite 
narrow distribution ($\sim 0.15$ dex), with a median H$\alpha$ flux of $10^{-14.3\pm 0.15}$ 
$\rm erg~ cm^{-2}~sec^{-1}$, 
revealing
that H$\alpha3$ is a rather homogeneous survey, despite the fact that observations 
were spread over almost one decade. At the distance of Virgo, our typical sensitivity 
corresponds to an unobscured SFR level of $1.3 \times 10^{-3}$ $\rm M_\odot~yr^{-1}$ at $1\sigma$, 
computed as outlined in Sect. \ref{sfr_calib}.
Panel (b) shows the distribution of morphological types from the ALFALFA\ galaxy catalogue
(dashed line) and from H$\alpha3$. 
Perhaps not surprisingly, an HI selected sample is strongly biased 
towards spirals and irregular galaxies (Gavazzi et al. 2008) at the depth achieved 
by ALFALFA. Stellar properties for our sample (solid lines) and for ALFALFA\ galaxies (dashed lines)
are presented in panel (c). 
Owing to the correspondence between stellar masses and HI masses (e.g. Gavazzi et al. 2008), 
the stellar distribution resembles the one for the gas masses, with a 
significant completeness down to less than $10^8$ M$_\odot$ (see Sect. \ref{compl}).
Galaxies in H$\alpha3$ span a wide range in color and gas fraction, 
allowing a statistical analysis of the star formation over a large space of 
parameters. In the color-magnitude diagram (d), HI selected galaxies lie almost exclusively 
in the blue cloud, while the red sequence (represented in the figure by the linear 
regression $g-i=-0.0585 \cdot (M_i+16) + 0.98$; Gavazzi et al. 2010) is grossly undersampled, 
as evident in the color magnitude diagram (Figure 7) of Haynes et al. (2011).
To detect the low level of atomic gas present in galaxies located in the green valley 
or even in the red sequence, deeper HI observations are required (e.g. Catinella et al. 2010). 
More detailed discussions of stellar and star formation properties
of the ALFALFA population overall are presented in Huang et al. (2011a) and Huang et al (2011b).

\section{Observations}\label{observ} 

Observations of HI selected galaxies 
from ALFALFA\ were completed in four runs of 9 nights each,
allocated from 2006 to 2009 by the Mexican Observatorio Astron\'omico Nacional (OAN)
at the San Pedro Martir Observatory (SPM, Baja California, Mexico).
Owing to the excellent weather conditions which are frequently encountered at 
SPM in the late Spring, 
we were able to observe mostly in photometric conditions:
8/9 nights in 2006, 9/9 in 2007-2008 and 5/9 in 
2009
During these runs\footnote {During an unfortunate run in 2010 the SIT3 CCD broke and was substituted with 
an outdated Thompson detector, badly affected by fringing in the red and with a low quantum efficiency. Due to the poor weather 
we could observe only 2 additional galaxies that are listed at the end of the Tables.}, we focused on the field surrounding the Virgo cluster, 
since most of the ALFALFA\ sources in Virgo  were already observed as part 
of a survey of optically selected galaxies started in 1999, using 
various telescopes: the  OHP and Calar Alto 1.2m (Boselli \& Gavazzi  2002),
the INT and NOT 2.5m (Boselli et al. 2002), the ESO 3.6m (Gavazzi et al. 2006) and the
SPM 2.1m (Gavazzi et al.2002a, b, 2006). 
We point to these papers for a detailed description of the observing strategies, 
data reduction and values for H$\alpha$ fluxes in this subsample.

\begin{figure}[!t]
\centering
\includegraphics[width=8cm,height=8cm]{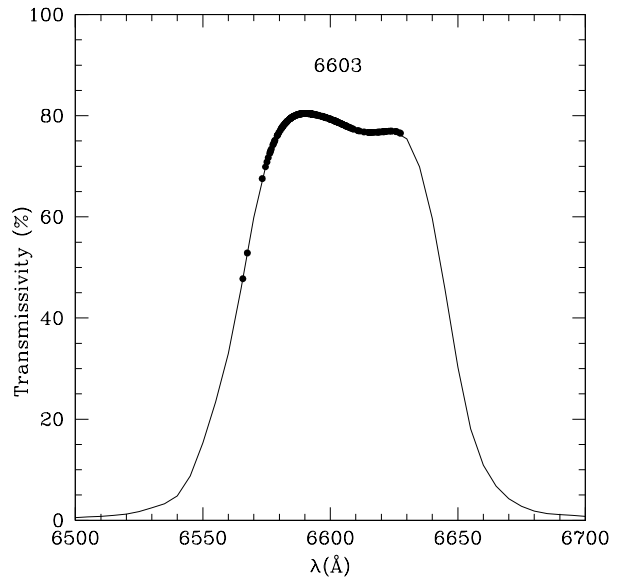}
 \caption{The transmissivity of the ON-band (6603 \AA~) filter. Filled circles mark the transmissivity
for $H\alpha$ at the redshift of the target galaxies. 
Two galaxies with $cz$=132 and 213 $\rm km~s^{-1}$ have been observed on the
steep shoulder of the filter transmission curve. They will not be further considered in the analysis.}
\label{filter}
\end{figure}

\begin{figure}[!t]
\centering
\includegraphics[width=8cm,height=8cm]{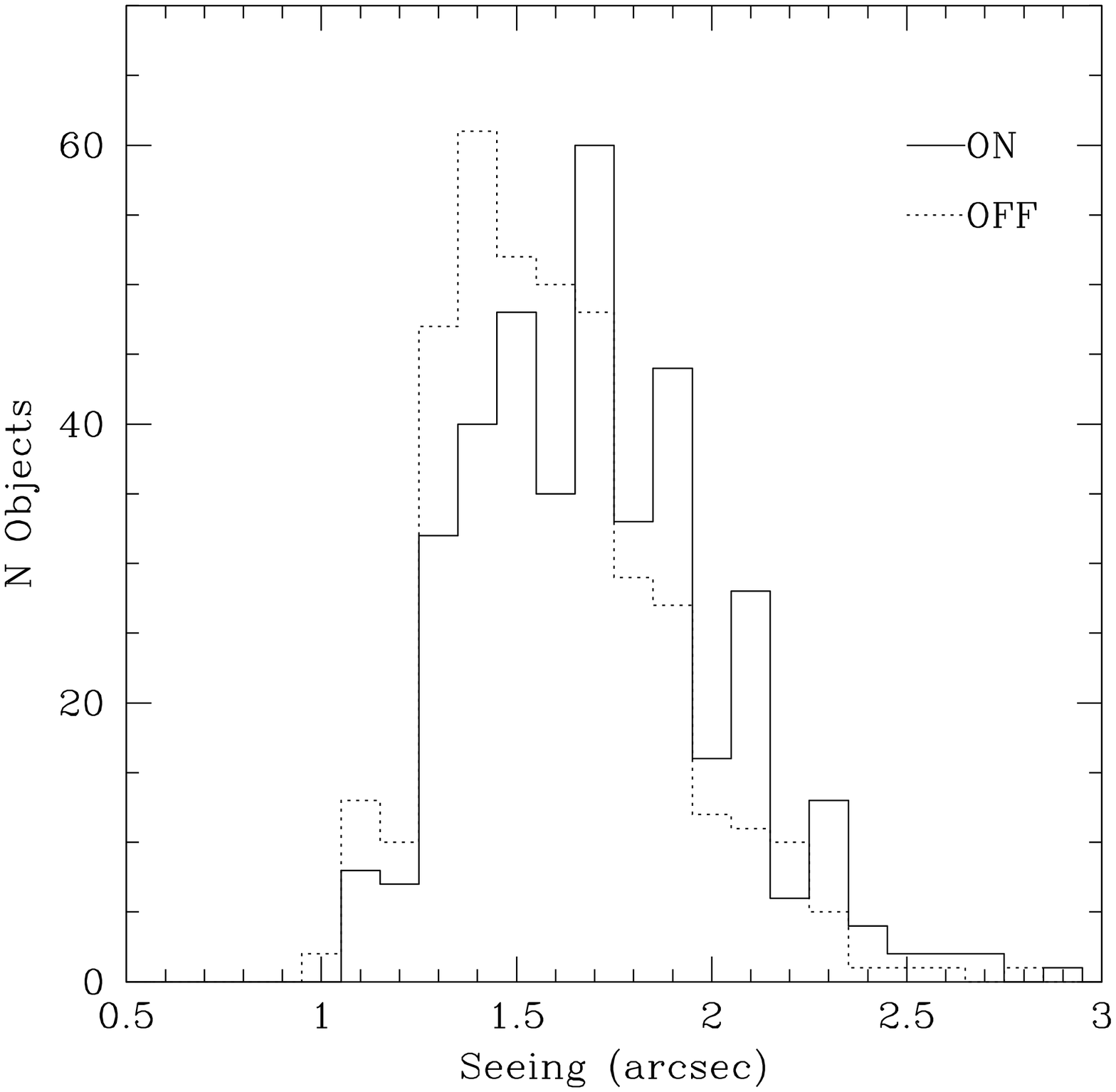}
 \caption{Point source FWHM measured on the final ON-band images (solid histogram) and OFF-band images (dashed histogram).
Poor telescope guiding performance and dome seeing affect the image quality, making the distribution
of the seeing slightly better in the shorter OFF exposures.}
\label{seeing}
\end{figure}

As for the data acquired between 2006 and 2009, we obtained narrow-band imaging in the 
H$\alpha$ emission line (rest frame $\lambda$ = 6562.8 \AA) with the $(f/7.5)$ Cassegrain 
focus imaging camera 
of the SPM 2.1 m telescope,  equipped with a SIT3 1024$\times$1024 pixels CCD detector with a 
pixel size of $0.31''$. The detector was used in the 1 e$^-$/ADU gain mode.
The redshifted H$\alpha$ line (ON-band frame) was imaged through a narrow band 
($\lambda$ 6603 \AA, $\Delta\lambda\sim 73$ \AA) interferometric filter, whose bandpass 
include also the [NII] lines. Except for two galaxies at lower velocities 
($cz<300$ $\rm km~s^{-1}$), this filter maximizes the throughput at the galaxy redshift, 
as shown in Fig. \ref{filter}\footnote{Two galaxies AGC4880 and AGC190160 with $cz$=4971 and 4954
$\rm km~s^{-1}$ respectively have been observed through a filter centered at $\lambda$ 6683 \AA.
The will not be considered in any further analysis}.
For each galaxy, we acquired multiple ON-band exposures with an integration time ranging 
from 15 to 60 min, according to seeing conditions and to source brightness.
The stellar continuum subtraction was secured by means of shorter (typically 3 to 5 min) 
observations taken through a broad-band ($\lambda$ 6231 \AA, $\Delta\lambda\sim 1200$ \AA) 
$r$-Gunn filter (OFF-band frames).
While the median seeing of the San Pedro Martir site is $\sim 0.''6$, 
the final FWHM for point sources in the images is affected by 
the poor telescope guiding and dome seeing.
For these reasons, the final distribution ranges from $1''$ to $\sim 2''.5$ (measured fitting a Gaussian
profile to the stars), 
with a mean seeing of $\sigma=1''.40 \pm 0''.28$  in the OFF-band images
and $\sigma=1''.6 \pm 0''.3$ in the longer ON-band exposures, as shown 
in Figure \ref{seeing}.

We derive the absolute flux levels with reference stars Feige34 and HZ44 from the catalogue of 
Massey et al. (1988), observed every $\sim$ 2 hours. Repeated measurements gave $<$ 5\% 
differences that we assume as the 1$\sigma$ photometric uncertainty. 
\footnote{ The stability of the photometry during each runs was such that we were able to detect a zero-point
decrease of 0.12 dex in 4 years due to loss of reflectivity of the mirrors.}
An handful number of galaxies have been imaged in transparent but not photometric 
conditions and for these objects we derive only 
the H$\alpha$ equivalent width (EW; insensitive to the absolute flux level), 
but not the H$\alpha$ flux.

We list information for individual galaxies in Table \ref{hextra}, as follows: 
\begin{itemize}
\item{Column  1:} AGC designation, from Haynes et al. (2011); 
\item{Column  2:} observing date (yy-mm-dd UT);
\item{Columns 3 and 4:} duration and number of individual ON-band exposures;
\item{Column  5:} average air mass during the ON-band exposures;
\item{Column  6:} adopted photometric zero point;
\item{Column  7:} FWHM of point sources (arcsec) in the ON-band frames, as measured on the images;
\item{Columns 8 and 9:} duration and number of individual OFF-band exposures;
\item{Column 10:} FWHM of point sources (arcsec) in the OFF-band frames, as measured on the images;
\item{Column 11:} normalization factor $n$ of the OFF-band frames (see next section).
\end{itemize}

\section {Data Reduction}
\label{data} 

\subsection {Image Analysis}
\label{analysis}  

We reduce the CCD frames following the procedure described Gavazzi et al. (2002b), 
using the STSDAS and GALPHOT  IRAF packages.
To compensate the spatial differences in the detector response, each image is bias subtracted 
and divided by the median of several flat-field exposures obtained during twilight 
in regions devoid of stars. 
When three exposures are available, we adopt a median combination of the realigned images 
to reject cosmic rays in the final stack. For galaxies observed in single exposures, we reject 
cosmic rays by direct inspection of the frames. For each frame, we subtract a mean local sky background, computed 
around the galaxy using the GALPHOT tasks MARKSKY and SKYFIT.
Over the typical spatial scale of galaxies ($50''-200''$) the mean background varies by $\sim 10\%$ 
of the sky rms per pixels. This is caused by residual patterns after flat-fielding
and represents the dominant source of error in low S/N regions. 
Over extended objects, the inability to subtract the sky with high accuracy
introduces an additional error on the final flux, which we properly account for in 
computing the error budget.

\subsection{Integral Photometry}

Due to the proximity of the two [NII] emission lines ($\lambda$ 6548-6584~\AA)
to the H$\alpha$ line, the flux measured in the ON-band observations
refers to a combination of H$\alpha$+[NII]. 
While a proper correction for [NII] emission is required before the final SFR is computed, 
in this section we will generically refer to H$\alpha$ as the total flux H$\alpha$+[NII].

Fluxes and EWs of the H$\alpha$ line can be recovered from narrow ON-band
observations by subtracting the stellar continuum contribution estimated using OFF-band images,
once these are normalized to account for the ratio of the transmissivity of the two filters
and the difference in exposure time.
For each galaxy, we derive the normalization coefficient $n$ by  
assuming that field stars have  no significant H$\alpha$ emission on average
and therefore they have identical continuum levels in the ON- and OFF-band frames
(see however Spector et al. 2011).

Once the normalization coefficient is known, we derive the 
integrated H$\alpha$ flux performing aperture photometry on both the OFF- and ON-band 
sky subtracted frames. First, we derive the integrated net counts $C_{NET}$ as:
\begin{equation}
{C_{NET} = C_{ON} - n\cdot C_{OFF}^{raw}=C_{ON} - C_{OFF}.}
\end{equation}
where we define the normalized OFF-band counts $C_{OFF}\equiv n\cdot C_{OFF}^{raw}$
with $C_{OFF}^{raw}$ the measured counts.
The net flux and EW in the H$\alpha$ line are then given by:
\begin{equation}
{F(H\alpha)_o = 10^{Zp} \frac{C_{NET}}{T~R_{ON}(H\alpha)}}
\end{equation}
\noindent
and 
\begin{equation}
{H\alpha EW_o = \frac{\int R_{ON}(\lambda)d\lambda}{R_{ON}(H\alpha)} \frac{C_{NET}}{n\cdot C_{OFF}^{raw}}}\:.
\label{EWo}
\end{equation}
\noindent
where $T$ is the integration time (sec), $10^{Zp}$ is the ON-band zero point ($\rm erg~ cm^{-2}~sec^{-1}$) 
corrected for atmospheric extinction 
and $R_{ON}(\lambda)$ is the transmissivity of the ON-filter at the wavelength of the redshifted $H\alpha$ line.
Finally, since the stellar continuum is estimated using a broad band $r$ filter that 
includes the H$\alpha$ line,  a non-negligible ($\sim$ 10\%) correction must be included 
(see Boselli et al 2002 and Gavazzi et al. 2006): 
\begin{equation}
{F(H\alpha) = F(H\alpha)_o \times \left(1+{\frac{\int R_{ON}(\lambda)d\lambda}{\int R_{OFF}(\lambda)d\lambda}}\right)}
\label{flux}
\end{equation}
and
\begin{equation}
{H\alpha EW=H\alpha EW_o \times \frac{\left(1+
\frac{\int R_{ON}(\lambda)d\lambda}{\int R_{OFF}(\lambda)d\lambda}\right)}
{\left(1 - 
\frac{H\alpha EW_o}{\int R_{OFF}(\lambda)d\lambda}\right)}}\:,
\label{EW}
\end{equation}
where $R_{OFF}(\lambda)$ is the transmissivity of the OFF filter.

\subsection{The SFR Calibration}
\label{sfr_calib}

\begin{figure}[-t]
\centering
\includegraphics[width=8cm,height=8cm]{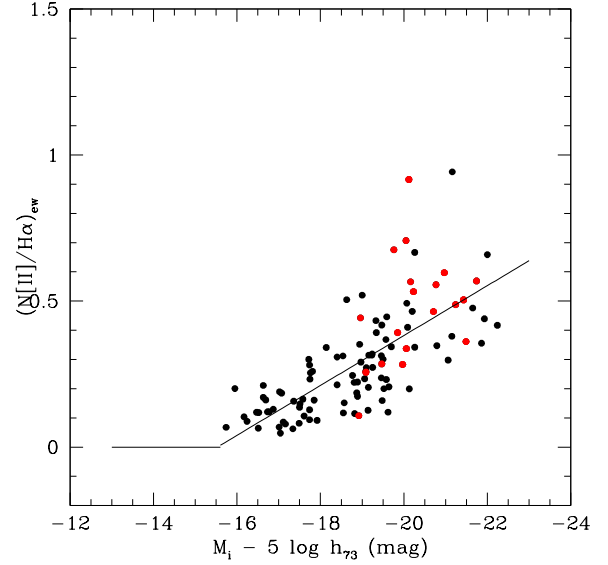}
 \caption{The ratio $([NII]/H\alpha)_{ew}$ derived from drift-scan spectra as a function of $M_i$,
 exhibiting the variation expected for the mass-metallicity relation. Red points 
 mark AGNs. The line indicates the linear fit to the data adopted when drift-scan spectra are unavailable.
 }
\label{deb_graph}
\end{figure}

The star formation rate is derived from the observed, integrated H$\alpha$ flux ($F(H\alpha)$) 
after the following corrections are applied:
i) Galactic extinction, ii) deblending from [NII], iii) internal 
extinction (Boselli at al. 2009).

i)  Corrections for Galactic extinction $A$ are computed using the color excess $E(B-V)$
obtained from the far-IR dust map of Schlegel et al., (1998). For the broad band photometry, 
we assume $A(R)=2.3E(B-V)$ and $A(I)=1.5E(B-V)$ (Cardelli et al, 1989), while for the $H\alpha$ fluxes 
we use $A(H\alpha)=0.6A(B)=2.6E(B-V)$  (Kennicutt et al, 2008, Cardelli et al, 1989).

ii) The correction for [NII] deblending is obtained by fitting the ratio $([NII]/H\alpha)_{ew}$ vs 
absolute $i$-band relation. This requires that AGNs (Seyfer+LINERS) are first identified (and disregarded)
using the nuclear (3 arcesc) SDSS spectra and the BTP (Baldwin et al. 1981) diagnostic. 
For doing so  the Balmer lines are corrected for underlying absorption by 5 \AA~  for $H\beta$   
(Kennicutt, 1992, Gavazzi et al, 2004) and by 1.3 \AA~ for $H\alpha$ (Gavazzi et al, 2011).  
We define as Seyfert those galaxies that have the ratio of $([NII]/H\alpha)_{ew}>0.5$ and  $([OIII]/H\beta)_{ew} > 3$
and LINERS galaxies that have the ratio of $([NII]/H\alpha)_{ew}>0.5$ and  $([OIII]/H\beta)_{ew} \leq 3$.
After excluding both classes of AGNs, we perform a linear fit between the ratio $([NII]/H\alpha)_{ew}$ 
and the absolute 
$i$-band magnitude (see Figure \ref{deb_graph}), corresponding to a well established 
mass-metallicity relation (Tremonti et al. 2004). We obtain a reliable fit 
with $([NII]/H\alpha)_{ew}=-0.854\times M_i-1.326$.\\ 
The corrected flux is:\\
 $F(H\alpha_{MW;DB}) = F(H\alpha_{MW})/(1+(1.34 \times [NII]/([NII]/H\alpha)_{ew}))$ \\
where the measured $([NII]/H\alpha)_{ew}$ is used when the a drift-scan spectrum is available from GOLDMine, otherwise
the ratio is obtained from the fit with $M_i$.

iii)  The correction for internal extinction is performed using $A_{Ha}$ from the Balmer decrement
if integrated drift-scan mode spectra are available and $H\alpha$ and $H\beta$ are both detected in emission
(see an asterisk in Column 9 of Table \ref{haspec}). 
Alternatively, when integrated drift-scan mode spectra are unavailable one can apply  
an average correction as function of the B band luminosity as proposed by  Lee et al. (2009):
$A_{Ha} =  1.971+0.323 \times B+0.0134 \times B^2 $ for  $B>-14$, otherwise $A_{Ha}=0.10$.
Mind however that such dependence is very poorly defined, thus we prefer to give in Table \ref{haspec} both  values
(with and without the  Lee et al. 2009 correction) and let the reader to decide which value to adopt.
To obtain such a correction one must first
convert SDSS g(AB) magnitudes into Johnson B mag, adopting the relation:
	$B =  g \times 0.983 +0.692 ~\rm mag $ (taken from GOLDMine). 
	In this case the dust extinction corrected flux becames:\\					
        $F(H\alpha_{MW;DB;AA}) = F(H\alpha_{MW;DB}) - (A_{Ha}/-2.5)$ \\ 
	Finally we derive the corrected $H\alpha$ luminosity:\\
	$ L(H\alpha_{MW;DB;AA}) = F(H\alpha_{MW;DB;AA}) + 48+ \log((3.086^2)\times 4 \pi \times D^2)$ \\	
	where $D$ is the distance in Mpc.  The star formation rate:		   
	log $(SFR) =  L(H\alpha_{MW;DB;AA}) - 41.1024 $ according to Kennicutt (1998).\\

The results of integral photometry as derived from the present observations are 
listed in Table \ref{haspec} as follows:

\begin{itemize}

\item{Column 1: AGC designation, from Haynes et al. (2011).} 
\item{Columns 2 and 3: R.A. and Dec. (J2000)}
\item{Column 4: equivalent width of H$\alpha$ + [NII] (in \AA) as given in equation (\ref{EW})} 
\item{Column 5: 1$\sigma$ uncertainty on the H$\alpha$ + [NII] equivalent width as given in equation (\ref{sigmaEW})} 
\item{Column 6: log of H$\alpha$ + [NII] (in $\rm erg~cm^{-2}~s^{-1}$) flux as given in equation (\ref{flux})}
\item{Column 7: log of 1$\sigma$ uncertainty on the H$\alpha$ + [NII] flux as given in equation (\ref{sigmaflux})}
\item{Column 8: log of SFR obtained in Sect. \ref{sfr_calib}, without correction for internal extinction.}
\item{Column 9: log of SFR\footnote{
Among galaxies detected in photometric conditions, SFR is given only for objects strictly
belonging to the H$\alpha3$ sample, i.e. in the interval 
$\rm 11^h < R.A. <16^h\, ; \,4^o< Dec. <16^o$; and for $350<cz<2000$ $\rm km ~s^{-1}$ (outside Virgo)
and $350<cz<3000$ $\rm km ~s^{-1}$ (inside Virgo).
For few sporadic galaxies observed in spite of not belonging to such interval,
we give flux and EW, but we don't convert it into SFR.} including the correction for internal extinction 
using the Balmer decrement when a drift-scan spectrum is available (see *) or as
proposed by Lee et al. (2009) in $\rm M_\odot~yr^{-1}$}.
\item{Column 10: sky quality: P = photometric ($\sigma < 5\%$), T = transparent ($5\%<\sigma<10\%$)}
\item{Column 11: Atlas figure}
\end{itemize}
 
\noindent We cross correlated our catalogue with the 11 Mpc volume H$\alpha$ survey by Kennicutt et al. (2008) and
we found 5 galaxies in common (see Table \ref{our_ken}). In spite of the paucity of the data, the agreement
between the two sets of measurements is satisfactory.

\scriptsize
\tiny
\begin{table}[h!]
\caption{Comparison between the photometry from this work and from Kennicutt et al. 2008
}
\begin{tabular}{lcccc}
\hline
AGC &         $\rm \log F (H\alpha+[NII])$           & $\rm EW$  &     $\rm \log F_{Ken}$       & $\rm EW_{Ken}$  \\
    &  $\rm erg~cm^{-2}~s^{-1}$      &  $\AA$    &     $\rm erg~cm^{-2}~s^{-1}$ & $\AA$           \\
(1) &            (2)             &          (3)              &   (4)       &     (5)              \\  
\hline
5456 &    -12.28 $\pm$ 0.04  &  42.33 $\pm$ 4.23  & -12.32 $\pm$ 0.01  &  46 $\pm$ 6 \\
6082 & $<$-12.13 $\pm$ 0.00  &   0.00 $\pm$ 3.20  & -12.92 $\pm$ 0.23  &  1  $\pm$ 1 \\
6272 &    -11.86 $\pm$ 0.16  &   9.11 $\pm$ 3.40  & -11.71 $\pm$ 0.03  &  14 $\pm$ 1 \\
6328 &    -11.26 $\pm$ 0.14  &   9.78 $\pm$ 3.16  & -11.53 $\pm$ 0.04  &  5  $\pm$ 1 \\
8091 &    -12.40 $\pm$ 0.04  &  90.03 $\pm$ 10.69 & -12.35 $\pm$ 0.01  &  103 $\pm$ 9 \\
\hline
\end{tabular}
\label{our_ken}
\end{table}
\normalsize

\begin{figure}[-t]
    \centering
    \includegraphics[width=8cm,height=8cm]{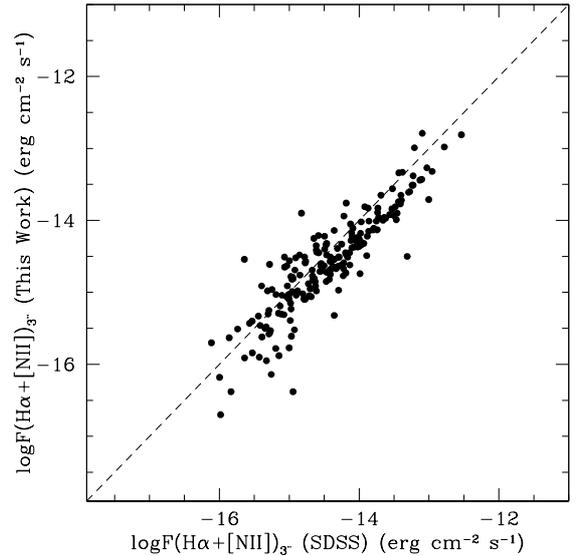}
    \caption{Comparison between the H$\alpha+[NII]$ flux extracted in 3 arcsec aperture centered on the nucleus (from this work) 
    and the flux in the
    H$\alpha+[NII]$  lines given by SDSS in 3 arcsec fibre spectrum (for detections with SN$>2$). The dashed line gives the one-to-one relation.
    }
    \label{3arcsec}
    \end{figure}

An independent check of the calibration of our H$\alpha$ measurements has been performed on a significant number of detections
by comparing the fluxes determined in 3 arcsec 
nuclear apertures in our H$\alpha$ images with the flux in the H$\alpha+[NII]$  lines listed in the SDSS spectral database obtained 
in 3 arcsec fibres (after removing all measurements not obtained in the nuclear regions). The comparison, given in Figure \ref{3arcsec},
shows a very satisfactory agreeement between the imaging and the spectral flux determinations.

\begin{figure}
\centering
\includegraphics[width=8cm,height=8cm]{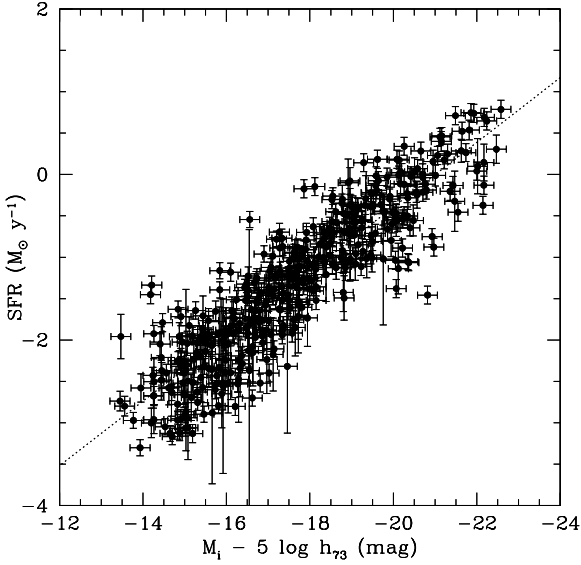}
 \caption{The corrected SFR derived from this work (see Column 9 of Table \ref{haspec})  
 as a function of the absolute $i$ band magnitude $M_i$}.
\label{SFR_i}
\end{figure}

The final SFRs, plotted against $M_i$  are presented in Figure \ref{SFR_i}. The error bars are obtained combining in quadrature 
the errors on $F(H\alpha)$
(see eq. 11) with the error on the coefficient of absorption from the Milky Way. 
Errors on the extinction coefficient $A_{Ha}$ and on the correction for 
deblending are not considered (see Boselli et al. 2009).
The linear regression is SFR=$-0.39*M_i-8.21$, with r=0.87, i.e. in the whole sample the global SFR is proportional to the
$i$ band luminosity.

\section{Summary and Future Prospects}
\label{summary}

     This is the first paper of a series devoted to
     H$\alpha3$, the H$\alpha$ narrow-band imaging survey of 
    galaxies carryed out with the San Pedro Martir 2.1m telescope (Mexico), 
    selected from the HI Arecibo Legacy Fast ALFA Survey (ALFALFA). 
    
    The first sample includes $\sim 400$  targets
    in the Local Supercluster for the sky region  
    $\rm 11^h < R.A. <16^h\, ; \,4^o< Dec. <16^o; 350< \rm cz<2000$ $\rm km ~s^{-1}$
    including the Virgo cluster.
    
    At the distance of Virgo (17 Mpc) and given the sensitivity of ALFALFA ~
    the targets selected for the H$\alpha$  follow-up contain more than $10^{7.7}$ M$_\odot$ of neutral atomic hydrogen.
    H$\alpha3$, complete for $M_{HI}>10^{8}$ M$_\odot$,
    provides the full census 
    of the star formation in HI rich galaxies of the local universe
    over a broad range of stellar masses, from dwarf galaxies with $10^{7.5}$ $M_\odot$ up to giants with $10^{11.5}$ $M_\odot$.
    Not unexpectedly, only an handful number of detections is identified with galaxies on the red sequence,
    while the majority are late-type, from giant spirals (Sa-Sd) to dwarf Irr-BCDs.
       
    In this paper, we present the properties of the H$\alpha$ galaxy sample, 
    together with H$\alpha$ fluxes and equivalent widths for the yet unpublished subsample observed between 2006 and 2009. 
     The integrated H$\alpha$ fluxes are corrected for galactic and internal extinction and for
    [NII] contamination  providing us with the global star formation rates (SFR). 
     Given the sensitivity of the present H$\alpha$ ~observations, 
     we detect galaxies with an 
     unobscured SFR density above $3x10^{-9}$ M$_\odot$ yr$^{-1}$ pc$^{-2}$  at $1\sigma$.

    The analysis of the integrated  quantities (global SFR) produced by H$\alpha3$ will be carryed out in Paper II of this series
    (Gavazzi et al. 2012).
    By using hydrogen recombination lines as a tracer of recent star formation, we aim to 
      investigate the relationships between atomic neutral gas and newly formed stars
      in different environments (cluster and field), morphological types 
      (spirals and dwarfs), and over a wide range of stellar masses 
      ($\sim 10^{7.5}-10^{11.5}$ M$_\odot$). 
  Paper III will contain the extension of H$\alpha3$ to the Coma supercluster
  ($\rm 10^h < R.A. <16^h\, ; \,24^o< Dec. <28^o; 3900<cz<9000$ $\rm km ~s^{-1}$).
  Being approximately 6 times more distant than Virgo, galaxies selected by ALFALFA at Coma 
  contain about 35 times more HI than those at Virgo.
  Hence ALFALFA will be complete for $\geq 10^{9.5}$ M$_\odot$, thus for giant galaxies.
  The cost of missing completely the population of dwarf galaxies will
  be compensated  by the fact that
  at $cz>5000$ the shredding problem affecting the SDSS completeness is much less severe than at Virgo,
  hence making it possible to extract a catalogue of optically selected candidates from the SDSS database. 
  This will allow us to investigate in details the differences between the optical  
  and the radio selection functions.

\appendix
\section{The error budget}
\label{errors}

As mentioned in Sect. \ref{analysis}, for extended sources, the dominant source of error in the final H$\alpha$ fluxes
is associated with variations of the background on scales similar to the source, 
which depend on the quality of the flat-fielding. We measure the background in several regions comparable with the size
of the galaxies and we establish that this fluctuation is on average $\sim 10\%$ of 
the sky $rms$ on the individual pixels. This error is dominant over the Poisson statistical uncertainty on the number counts. 
Therefore, the total uncertainty on the ON and OFF
counts is proportional to the number of pixels $N_{pixel}$ occupied by each galaxy,
as derived from the optical major and minor axes, $a$ and $b$ respectively (see Gavazzi et al. 2002a):

\begin{eqnarray}
\sigma_{ON} & = & 0.1 ~rms_{ON}~ N_{pixel} \\
 & & \nonumber \\
\sigma_{OFF}^{raw} & = & 0.1 ~rms_{OFF}~ N_{pixel}
\end{eqnarray}

An additional source of error affecting the OFF counts derives from the uncertainty on the
normalization coefficient $n$ which we estimate to be $\sim$ 3\%, thus:

\begin{equation}
\sigma_{OFF} = \sqrt{(n\cdot \sigma_{OFF}^{raw})^2 + (0.03~n\cdot C_{OFF}^{raw})^2.}
\end{equation} 

\noindent The error on the NET counts is defined as:

\begin{equation}
\sigma_{NET} = \sqrt{(\sigma_{ON})^2 + (\sigma_{OFF})^2} .
\end{equation}

\noindent
The error on the H$\alpha$ flux finally becomes:
\begin{equation}
\sigma_{F_o}  =  \left[\left(\frac{10^{Zp}\sigma_{C_{NET}}} {T R_{ON}(H\alpha)} \right)^2 + 
\left(0.05 \cdot 10^{Zp} \ln\,(10)\, F_o\right)^2 \right]^{0.5}
\label{sigma_fo}
\end{equation}

\begin{equation}
{\sigma_{F} = \sigma_{F_o} \times
\left(1+{\frac{\int R_{ON}(\lambda)d\lambda}{\int R_{OFF}(\lambda)d\lambda}}\right)} .
\label{sigmaflux}
\end{equation}

\noindent The second term in Eq. (\ref{sigma_fo}) accounts for 
the uncertainty on the photometric calibration, which we estimate to be 5\%.

Similarly, for the EWs, we compute the final error as:

\begin{equation}
{\sigma_{EW_o} = \frac{\int R_{ON}(\lambda)d\lambda}{R_{ON}(H\alpha)} 
\left[\left(\frac{\sigma_{NET}}{n\cdot C_{OFF}^{raw}}\right)^2 + \, 
\left(\frac{C_{NET}\sigma_{OFF}}{(n\cdot C_{OFF}^{raw})^2}\right)^2\right]^{0.5}}
\end{equation}

\begin{equation}
{
\sigma_{EW} = \sigma_{EW_o} \times 
\frac{\left(1+
\frac{\int R_{ON}(\lambda)d\lambda}{\int R_{OFF}(\lambda)d\lambda}\right)}
{\left(1 - 
\frac{H\alpha EW}{\int R_{OFF}(\lambda)d\lambda}\right)^2} .
}
\label{sigmaEW}
\end{equation}

In conclusion this error budget results from several components (photometric accuracy, flat fielding),
even if a systematic uncertainty on the normalization factor, as derived from the measurement of foreground
stars, is the dominant source of error (Spector et al. 2011).

\section{The Atlas}\label{atlas}

Images of the OFF and NET frames of galaxies with  H$\alpha$ observations presented in this work
are given separately for
102 galaxies with substantial H$\alpha$ structure in Fig. 9; for
84 marginal detections ($<$2$\sigma$) or with unresolved/complex H$\alpha$ emission in Fig. 10. 
For the  47 remaining galaxies with no H$\alpha$ emission,  the OFF-band images
are shown in Fig. 11. Galaxies are labeled with their celestial coordinates. A 1 arcmin bar is given.

\begin{acknowledgements}
This work is dedicated to the memory of Gaby Garcia who payed with his life the passion for his work. 
We thank the night operators, specially Felipe Montalvo and Salvador Monrroy for their collaboration, 
the resident astronomers at SPM for their assistance during the observations
and the mexican TAC for the generous time allocation.
We acknowledge useful discussions with Luis Aguillar, Luis Carrasco, Matteo Fossati, Michael Richter and Giulia Savorgnan.
We thank L. Giordano, D. Burlon, E. Farina, C. Pacifici and V. Presotto for their help
during the observations and L. Cortese and S. Zibetti, F. Martinelli and I. Arosio for their support in the data reduction.\\ 
The authors would like to acknowledge the work of the entire ALFALFA collaboration team
in observing, flagging, and extracting the catalog of galaxies used in this work.
This research has made use of the GOLDMine database (Gavazzi G. et al. 2003)
and of the NASA/IPAC Extragalactic Database (NED) which is operated 
by the Jet Propulsion Laboratory, California Institute of Technology, under contract with the
National Aeronautics and Space Administration. \\
Funding for the Sloan Digital Sky Survey (SDSS) and SDSS-II has been provided by the 
Alfred P. Sloan Foundation, the Participating Institutions, the National Science Foundation, 
the U.S. Department of Energy, the National Aeronautics and Space Administration, 
the Japanese Monbukagakusho, and 
the Max Planck Society, and the Higher Education Funding Council for England. 
The SDSS Web site is \emph{http://www.sdss.org/}.
The SDSS is managed by the Astrophysical Research Consortium (ARC) for the Participating Institutions. 
The Participating Institutions are the American Museum of Natural History, Astrophysical Institute Potsdam, 
University of Basel, University of Cambridge, Case Western Reserve University, The University of Chicago, 
Drexel University, Fermilab, the Institute for Advanced Study, the Japan Participation Group, 
The Johns Hopkins University, the Joint Institute for Nuclear Astrophysics, the Kavli Institute for 
Particle Astrophysics and Cosmology, the Korean Scientist Group, the Chinese Academy of Sciences (LAMOST), 
Los Alamos National Laboratory, the Max-Planck-Institute for Astronomy (MPIA), the Max-Planck-Institute 
for Astrophysics (MPA), New Mexico State University, Ohio State University, University of Pittsburgh, 
University of Portsmouth, Princeton University, the United States Naval Observatory, and the University 
of Washington.\\
G.G. acknowledges financial support from italian MIUR PRIN contract 200854ECE5.
R.G. and M.P.H. are supported by US NSF grants AST-0607007 and AST-1107390 and by a Brinson Foundation grant.
\end{acknowledgements}

\newpage

\newpage
\onecolumn
\scriptsize
\tiny
\begin{longtable}{rcccccccrccccc}
\caption{Basic data for the 235 target galaxies}\\
\hline
AGC & $R.A.~(J2000)$ & $DEC$  & CGCG & UGC & NGC & IC & Ty & $cz$            &     $a$	    & $b$        & $i$      & $g$  &  $Dist$	 \\
    &  hhmmss.s   & $^o~'~''$ &      &     &     &    &    & $\rm km~s^{-1}$ &  arcmin      & arcmin   &  mag  & mag  & Mpc \\
(1) &   (2)       &  (3)   & (4)  & (5) & (6) & (7)& (8)& (9)             &  (10)    &    (11)      &  (12)    & (13)    & (14)  \\
\hline                                                                                                                              
\endfirsthead
\multicolumn{14}{l}{\footnotesize\itshape
Continued from previous page} \\
\hline
AGC & $R.A.~(J2000)$ & $DEC$  & CGCG & UGC & NGC & IC & Ty & $cz$            &     $a$	    & $b$        & $i$      & $g$  &  $Dist$	 \\
    &  hhmmss.s   & $^o~'~''$ &      &     &     &    &    & $\rm km~s^{-1}$ &  arcmin      & arcmin   &  mag  & mag  & Mpc \\
(1) &   (2)       &  (3)   & (4)  & (5) & (6) & (7)& (8)& (9)             &  (10)    &    (11)      &  (12)    & (13)    & (14)  \\
\hline
\endhead
\hline
\multicolumn{12}{r}{\footnotesize\itshape
Continued in next page} \\
\endfoot
\hline
\multicolumn{12}{r}{\footnotesize\itshape
Concluded from previous page} \\
\endlastfoot
    4880  &   091517.0 &  115308  & 62010  & 4880  &  -   & 530  &   Sab   & 4971  &  1.8  & 0.6 & 12.6 & 13.8  &  66.5   \\
  190160  &   091647.8 &  112709  & 62015  &   -   &  -   &  -   &    -    & 4954  &  0.5  & 0.1 & 13.7 & 13.8  &  66.7   \\
  202488  &   100610.8 &  110602  &   -    &   -   &  -   &  -   &    -    & 2505  &  0.1  & 0.1 & 15.5 & 16.0  &  32.7   \\
    5456  &   100719.7 &  102143  & 64068  & 5456  &  -   &  -   &   Sm    & 537   &  1.6  & 0.8 & 13.1 & 13.5  &   5.9   \\
    5741  &   103442.8 &  111148  & 65059  & 5741  & 3279 & 622  &   Scd   & 1383  &  2.0  & 0.5 & 12.2 & 13.4  &  17.7    \\
  208399  &   104010.7 &  045432  &   -    &   -   &  -   &  -   &    -    &  747  &  0.5  & 0.3 & 15.7 & 16.9  &  20.0   \\
  202024  &   104457.5 &  115458  &   -    &   -   &  -   &  -   &   dS0   & 869   &  0.3  & 0.1 & 17.2 & 17.6  &  10.6   \\
    6077  &   110002.4 &  145028  & 95085  & 6077  & 3485 &  -   &   Sb    & 1432  &  1.8  & 1.5 & 11.6 & 12.4  &  24.5   \\
\end{longtable}
\normalsize

\newpage

\scriptsize
\tiny
\begin{longtable}{rcrccccrccc}
\caption{$H\alpha$ observational specifications of the 235 target galaxies}\\
& & \multicolumn{1}{l}{ON}&\multicolumn{5}{r}{OFF}\\ 
\hline
AGC &    Date        & $T_{exp}$  & $N_{exp}$ & A.M. &  $Zp$                     & Seeing      & $T_{exp}$   & $N_{exp}$  &   Seeing    & $n$ \\
    &    yymmdd      &   sec      &           &      &  $\rm erg~cm^{-2}~s^{-1}$ &   arcsec    &  sec        &            &    arcsec   &       \\
(1) &      (2)       &  (3)       &    (4)    &  (5) &          (6)              &  (7)        &  (8)        & (9)        &  (10)       &  (11)   \\
\hline 
\endfirsthead
\multicolumn{11}{r}{\footnotesize\itshape
Continued from previous page} \\
\hline   
& & \multicolumn{1}{l}{ON}&\multicolumn{5}{r}{OFF}\\    
\hline  
AGC &    Date        & $T_{exp}$  & $N_{exp}$ & A.M. &  $Zp$                     & Seeing      & $T_{exp}$   & $N_{exp}$  &   Seeing    & $n$ \\
    &    yymmdd      &   sec      &           &      &  $\rm erg~cm^{-2}~s^{-1}$ &   arcsec    &  sec        &            &    arcsec   &       \\
(1) &      (2)       &  (3)       &    (4)    &  (5) &          (6)              &  (7)        &  (8)        & (9)        &  (10)       &  (11)   \\     
\hline 
\endhead
\hline  
\multicolumn{11}{r}{\footnotesize\itshape
Continued in next page} \\
\endfoot
\hline  
\multicolumn{11}{r}{\footnotesize\itshape
Concluded from previous page} \\
\endlastfoot                                                                                                               
    4880  &  2006-05-01     &	900  &  1  &   1.10   &  -15.52  &    1.6   &	     180   &	1   &	 1.8   &     0.376   \\  
  190160  &    2006-05-01   &	900  &  1  &   1.14   &  -15.52  &    2.0   &	     180   &	1   &	 1.7   &     0.464   \\  
  202488  &    2006-04-26   &	900  &  1  &   1.21   &  -15.51  &    1.8   &	     180   &	1   &	 1.8   &     0.486   \\  
    5456  &  2006-04-24     &	900  &  1  &   1.07   &  -15.51  &    1.8   &	     180   &	1   &	 1.5   &     0.460   \\  
    5741  &  2006-04-23     &	900  &  1  &   1.07   &  -15.51  &    2.2   &	     180   &	1   &	 1.9   &     0.460   \\  
  208399  &    2008-04-12   &  1200  &  3  &   1.13   &  -15.45  &    1.6   &	     300   &	3   &	 1.5   &     0.358   \\  
  202024  &    2006-04-25   &	900  &  3  &   1.08   &  -15.51  &    2.0   &	     240   &	3   &	 2.0   &     0.386   \\  
    6077  &  2007-04-19     &	900  &  1  &   1.12   &  -15.45  &    1.6   &	     180   &	1   &	 1.6   &     0.470   \\
\end{longtable}
\normalsize

\newpage

\scriptsize
\tiny
\begin{longtable}{rrrrrrrrrcc}
\caption{Integrated $H\alpha$ photometric parameters of the 235 target galaxies}\\
\hline
AGC &  $R.A.~(J2000)$ & $DEC$      &   $\rm H\alpha EW$  &  $\rm \sigma_{EW}$  &  $\rm \log~F(H\alpha)$   &  $\rm \log~\sigma_F$ &        $\rm \log~SFR$  &        $\rm \log~SFR$  	   & Quality & Fig.  \\
    &   hhmmss    & $^o~'~''$ &       $\AA$         &      $\AA$          &  $\rm erg~cm^{-2}~s^{-1}$	      &  $\rm erg~cm^{-2}~s^{-1}$ &         $\rm M_\odot~y^{-1}$	&         $\rm M_\odot~y^{-1}$	   &	     &       \\
(1) &   (2)       &      (3)  &        (4)          &           (5)       &  	       (6)		      &   (7)			  &        (8)  &      (9)     & (10)	 & (11)   \\
\hline   
\endfirsthead
\multicolumn{11}{r}{\footnotesize\itshape 
Continued from previous page} \\
\hline   
AGC &  $R.A.~(J2000)$ & $DEC$      &   $\rm H\alpha EW$  &  $\rm \sigma_{EW}$  &  $\rm \log~F(H\alpha)$   &  $\rm \log~\sigma_F$ &        $\rm \log~SFR$  &        $\rm \log~SFR$  	   & Quality & Fig.  \\
    &   hhmmss    & $^o~'~''$ &       $\AA$         &      $\AA$          &  $\rm erg~cm^{-2}~s^{-1}$	      &  $\rm erg~cm^{-2}~s^{-1}$ &         $\rm M_\odot~y^{-1}$	&         $\rm M_\odot~y^{-1}$	   &	     &       \\
(1) &   (2)       &      (3)  &        (4)          &           (5)       &  	       (6)		      &   (7)			  &        (8)  &      (9)     & (10)	 & (11)   \\
\hline 
\endhead
\hline
\multicolumn{11}{r}{\footnotesize\itshape
Continued in next page} \\
\endfoot  
\hline
\multicolumn{11}{r}{\footnotesize\itshape
Concluded from previous page} \\
\endlastfoot                                                                                                                        
   4880  &  091517.0 & 115308  &   15.93  &  3.43   &      -12.65 &   0.09 &	    -	 &	  -     & P &	9   \\     
 190160  &  091647.8 & 112709  &   29.31  &  3.54   &      -13.41 &   0.05 &	    -	 &	  -     & P &	9   \\     
 202488  &  100610.8 & 110602  &   33.73  &  4.01   &      -13.36 &   0.05 &	    -	 &	  -     & P &  10   \\     
   5456  &  100719.7 & 102143  &   42.33  &  3.62   &      -12.28 &   0.03 &	    -	 &	  -     & P &	9   \\     
   5741  &  103442.8 & 111148  &   16.60  &  3.20   &      -12.44 &   0.08 &	    -	 &	  -     & P &	9   \\     
 208399  &  104010.7 & 045432  &    0.00  &  3.89   & $<$  -14.42 &    -   &	    -	 &	  -     & P &  11   \\     
 202024  &  104457.5 & 115458  &    0.00  &  6.24   & $<$  -14.64 &    -   &	    -	 &	  -     & P &  11   \\     
   6077  &  110002.4 & 145028  &   16.99  &  4.08   &      -12.14 &   0.10 &	 -0.56   &    -0.28     & P &	9   \\
\end{longtable}
\normalsize

\end{document}